\documentclass[twocolumn,showpacs,preprintnumbers,amsmath,amssymb,prl,superscriptaddress]{revtex4}

\usepackage{bm}
\usepackage{graphicx}
\usepackage{color}

\newcommand{\bra}[1]{\langle #1|}
\newcommand{\ket}[1]{|#1\rangle}
\newcommand{\braket}[1]{\langle #1 \rangle}

\def\ee{\mathrm{e}}
\def\ii{\mathrm{i}}

\def\wp{\omega_{\text{p}}}
\def\DE{\Delta E}

\def\oH{\hat{H}}
\def\oa{\hat{a}}
\def\oad{\hat{a}^{\dagger}}

\begin{document}


\title{On the origin of strong photon antibunching in weakly nonlinear
photonic molecules}

\author{Motoaki Bamba}
\altaffiliation{E-mail: motoaki.bamba@univ-paris-diderot.fr}
\affiliation{Laboratoire Mat\'eriaux et Ph\'enom\`enes Quantiques,
Universit\'e Paris Diderot-Paris 7 et CNRS, \\ B\^atiment Condorcet, 10 rue
Alice Domon et L\'eonie Duquet, 75205 Paris Cedex 13, France}
\author{Atac Imamo\u{g}lu}
\affiliation{Institute of Quantum Electronics, ETH Z\"urich, 8093
Z\"urich, Switzerland}
\author{Iacopo Carusotto}
\affiliation{INO-CNR BEC Center and Dipartimento di Fisica, Universit\`a
di Trento, I-38123 Povo, Italy}
\author{Cristiano Ciuti}
\altaffiliation{E-mail: cristiano.ciuti@univ-paris-diderot.fr}
\affiliation{Laboratoire Mat\'eriaux et Ph\'enom\`enes Quantiques,
Universit\'e Paris Diderot-Paris 7 et CNRS, \\ B\^atiment Condorcet, 10 rue
Alice Domon et L\'eonie Duquet, 75205 Paris Cedex 13, France}

\date{\today}

\begin{abstract}
In a recent  work [T.~C.~H.~Liew and V.~Savona, Phys.~Rev.~Lett.~{\bf104}, 183601 (2010)] it was numerically shown that in a photonic
'molecule' consisting of two coupled cavities, near-resonant
coherent excitation could give rise to strong photon antibunching
with a surprisingly weak nonlinearity. Here, we show that a subtle
quantum interference effect is responsible for the predicted
efficient photon blockade effect. We analytically determine the
optimal on-site nonlinearity and frequency detuning between the pump
field and the cavity mode. We also highlight the limitations of the
proposal and its potential applications in demonstration of strongly
correlated photonic systems in arrays of weakly nonlinear cavities.
\end{abstract}


\maketitle The photon blockade is a quantum optical effect
preventing the resonant injection of more than one photon into a
nonlinear cavity mode \cite{Imamoglu1997PRL}, leading to antibunched
(sub-Poissonian) single-photon statistics. Signatures of photon
blockade have been observed by resonant laser excitation of an
optical cavity containing either a single atom \cite{Birnbaum2005N}
or a single quantum dot \cite{Faraon2008NP} in the strong coupling
regime. Arguably, the most convincing realization was based on a
single atom coupled to a micro-toroidal cavity in the Purcell regime
\cite{Dayan2008S}, suggesting that the strong coupling regime of
cavity-QED need not be a requirement. Concurrently, on the theory
side there has been a number of proposals investigating strongly
correlated photons in coupled cavity arrays \cite{Hartmann2006NP,
Greentree2006NP, Angelakis2007PRA} or one-dimensional optical waveguides
\cite{Chang2008NP}. The specific proposals based on the photon
blockade effect include the fermionization of photons in
one-dimensional cavity of arrays \cite{Carusotto2009PRL}, the
crystallization of polaritons in coupled array of cavities
\cite{Hartmann2010PRL}, and the quantum-optical Josephson
interferometer in a coupled photonic mode system
\cite{Gerace2009NP}.

\begin{figure}[tbp] 
\includegraphics[width=\hsize]{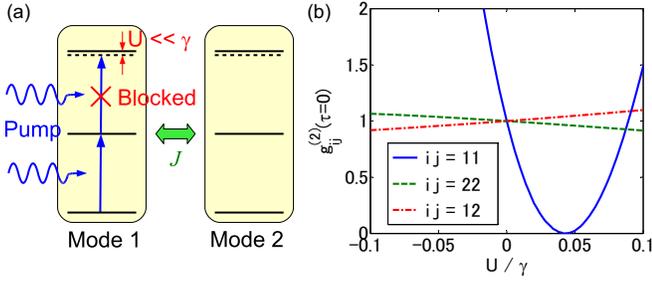}
\caption{(1) Sketch of the two coupled photonic modes.
The coupling strength is $J$, and the antibunching is obtained with a small nonlinear energy $U$
compared to mode broadening $\gamma$.
(b) Equal-time second-order correlation functions $g^{(2)}_{ij}(\tau=0)$ are plotted
as functions of nonlinearity $U = U_1 = U_2$ normalized to $\gamma$.
The nearly perfect antibunching is obtained at the pumped mode [$g^{(2)}_{11}(\tau=0) \simeq 0$]
for $U = 0.0428\gamma$. Parameters:
$\gamma_1 = \gamma_2 = \gamma$, $J = 3\gamma$, $E_1 = E_2 = \hbar\wp + 0.275\gamma$,
and $F_1 = 0.01\gamma$.}
\label{fig:1}
\end{figure}

It is commonly believed that photon blockade necessarily requires a
strong on-site nonlinearity $U$ for a photonic mode, whose magnitude
should well exceed the mode broadening $\gamma$. However, in a
recent work \cite{Liew2010PRL} Liew and Savona numerically showed
that a strong antibunching  can be obtained  with a surprisingly
weak nonlinearity ($U \ll \gamma$) in a system consisting of two
coupled zero-dimensional (0D) photonic cavities (boxes), as shown in
Fig.~\ref{fig:1}(a) \cite{Liew2010PRL}. Such a configuration can be
obtained, e.g., by considering two modes in two photonic boxes
coupled with a finite mode overlap due to leaky mirrors: the
corresponding tunnel strength will be designated with $J$. In Ref.~\cite{Liew2010PRL}
numerical evidence indicated that a nearly
perfect antibunching can be achieved for an optimal value of the
on-site repulsion energy $U$ and for an optimal value of the
detuning between the pump and mode frequency. However, a physical
understanding of the mechanism leading to strong photon antibunching
is needed to identify the limitations of the scheme in the context
of proposed experiments on strongly correlated photons, as well as
to determine the dependence of the optimal coupling and detuning on
the relevant physical parameters $J$ and $\gamma$.

In this letter, we show analytically that the surprising
antibunching effect is the result of a subtle destructive quantum
interference effect which ensures that the probability amplitude to
have two photons in the driven cavity is zero. We show that the weak
nonlinearity is required only for the auxiliary cavity that is not
laser driven and whose output is not monitored, indicating that
photon antibunching is obtained for a driven linear cavity that
tunnel couples to a weakly nonlinear one. We determine the
analytical expressions for the optimal coupling $U$ and for the pump
frequency detuning required to have a perfect antibunching as a
function of the mode coupling $J$ and broadening $\gamma$. Our
analytical results are in excellent agreement with fully numerical
solutions of the master equation for the considered system. Before
concluding, we discuss the experimental realization of such a scheme
by using cavities embedding weakly coupled quantum dots.
Moreover, we consider also the case of a ring of coupled photonic molecules
showing that strong antibunching persists in presence of intersite photonic correlations.

We consider two photonic modes coupled with strength $J$; each mode
has energy $E_j$ and an on-site photon-photon interaction strength
$U_i$ ($i = 1, 2$). The Hamiltonian is written as
\begin{align}
\oH & = \sum_{i=1}^2 \left[ E_i \oad_i \oa_i + U_i \oad_i\oad_i\oa_i\oa_i\right]
+ J(\oad_1\oa_2 + \oad_2\oa_1)
\nonumber \\ & \quad
+ F \ee^{-\ii\wp t}\oad_1 + F^*\ee^{\ii\wp t}\oa_1,
\end{align}
where $\oa_i$ is the annihilation operator of a photon in $i$-th
mode, $F$ and $\wp$ are the pumping strength and frequency,
respectively. Following Ref.~\cite{Liew2010PRL}, we first calculate
the second-order correlation function
$g^{(2)}_{ij}(\tau) = \braket{\oad_i\oad_j(\tau)\oa_j(\tau)\oa_i} /
\braket{\oad_i\oa_i} \braket{\oad_j\oa_j}$ in the steady state using the master
equation in a basis of Fock states \cite{verger06}.  The results are
shown as functions of nonlinearity $U$ in Fig.~\ref{fig:1}(b). As
already demonstrated in Ref.~\cite{Liew2010PRL}, we can get a strong
antibunching of the pumped mode ($g^{(2)}_{11}(0) \simeq 0$) for an
unexpectedly small nonlinearity $U = 0.0428\gamma$.

In order to understand the origin of the strong antibunching, we use
the Ansatz
\begin{align}
\ket{\psi}
& = C_{00}\ket{00} + \ee^{-\ii\wp t}\left( C_{10}\ket{10} + C_{01}\ket{01}\right)
\nonumber \\ & \quad
+ \ee^{-\ii2\wp t}\left(C_{20}\ket{20} + C_{11}\ket{11} + C_{02}\ket{02}\right)
+ \ldots,
\end{align}
to calculate the steady-state of the coupled cavity system. Here,
$\ket{mn}$ represents the Fock state with $m$ particles in mode 1
and $n$ particles in mode 2. Under weak pumping conditions ($C_{00}
\gg C_{10}, C_{01} \gg C_{20}, C_{11}, C_{02}$), we can calculate
the coefficients $C_{mn}$ iteratively. For one-particle states, the
steady-state coefficients are determined by
\begin{subequations}
\begin{align}
&(\DE_1-\ii\gamma_1/2) C_{10} + J C_{01} + F C_{00} = 0, \\
&(\DE_2-\ii\gamma_2/2) C_{01} + J C_{10} = 0, \label{eq:C01+C10=0} 
\end{align}
\end{subequations}
where $\DE_j = E_j - \hbar\wp$ and we consider a damping with rate
$\gamma_j$ in each mode. Since we assume weak pumping, the
contribution from the higher states ($C_{20}$, $C_{11}$, and
$C_{02}$) to the steady-state values of $C_{10}, C_{01}$ is
negligible. From Eq.~\eqref{eq:C01+C10=0}, the amplitude of mode 2
can be written as
\begin{equation} \label{eq:C01=C10} 
C_{01} = - \frac{J}{\DE_2-\ii\gamma_2/2} C_{10}.
\end{equation}
indicating that for strong photon tunneling ($J \gg |\DE_2| ,
\gamma_2$), the probability of finding a photon in the auxiliary
cavity is much larger than the driven cavity.

\begin{figure}[tbp] 
\includegraphics[width=\hsize]{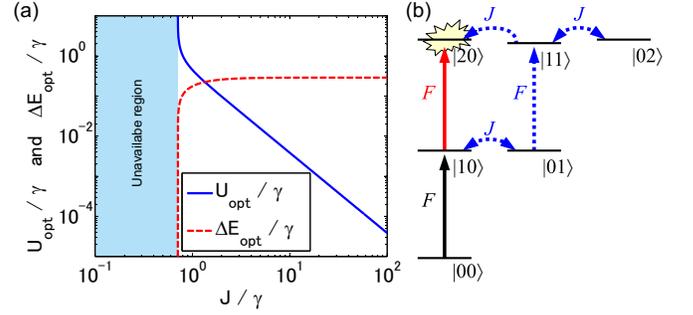}
\caption{(a) Optimal nonlinearity $U_{\text{opt}}$ and detuning $\Delta E_{\text{opt}}$ are plotted
as functions of coupling strength $J$ normalized to $\gamma$
($\gamma_1 = \gamma_2 = \gamma$ and $E_1 = E_2 = E$).
The perfect antibunching is obtained for $J > \gamma/\sqrt{2}$.
(b) Transition paths leading to the quantum interference responsible for the strong antibunching.
One path is the direct excitation from $\ket{10}$ to $\ket{20}$,
but it is forbidden by the interference with the other path drawn by dotted arrows.}
\label{fig:2}
\end{figure}

In the same manner, the coefficients of two-particle states are
determined by
\begin{subequations} \label{eq:two-particle} 
\begin{align}
&2(\DE_1+U_1-\ii\gamma_1/2) C_{20} + \sqrt{2} J C_{11} + \sqrt{2} F C_{10} = 0, \label{C20=C11+C10} \\
&(\DE_1+\DE_2-\ii\gamma_1/2-\ii\gamma_2/2) C_{11}
+ \sqrt{2} J C_{20} + \sqrt{2} J C_{02}
\nonumber \\ & \quad
 + F C_{01} = 0, \label{C20+C02+C01=0} \\
&2(\DE_2+U_2-\ii\gamma_2/2) C_{02} + \sqrt{2} J C_{11} = 0.
\end{align}
\end{subequations}
When we simply consider $E_1 = E_2 = E$, and $\gamma_1 = \gamma_2 = \gamma$,
the conditions to satisfy $C_{20} = 0$ are derived from Eqs.~\eqref{eq:C01=C10} and \eqref{eq:two-particle} as
\begin{subequations} \label{eq:for_C20=0} 
\begin{align}
&\gamma^2(3\DE+U_2) - 4\DE^2(\DE+U_2) = 2J^2U_2, \\
&12\DE^2 + 8\DE U_2 - \gamma^2 = 0.
\end{align}
\end{subequations}
For fixed $J$ and $\gamma$, from these equations, the optimal conditions (those that lead
to $C_{20} = 0$) are given by
\begin{subequations} \label{eq:U_DE_opt} 
\begin{align}
\DE_{\text{opt}} & = \pm \frac{1}{2} \sqrt{\sqrt{9J^4+8\gamma^2J^2} - \gamma^2 - 3J^2}, \\
U_{\text{opt}} & = \frac{\DE_{\text{opt}}(5\gamma^2+4\DE_{\text{opt}}{}^2)}{2(2J^2-\gamma^2)},
\end{align}
\end{subequations}
and, if $J \gg \gamma$, they are approximately written as
\begin{subequations} \label{eq:U_DE_opt_approx} 
\begin{align}
\DE_{\text{opt}} & \simeq \frac{\gamma}{2\sqrt{3}}, \\
U_{\text{opt}} & \simeq \frac{2}{3\sqrt{3}}\frac{\gamma^3}{J^2}. \label{eq:U_opt_approx} 
\end{align}
\end{subequations}
In Fig.~\ref{fig:2}(a), the optimal $\Delta E_{\text{opt}}$ and
$U_{\text{opt}}$ [Eq.~\eqref{eq:U_DE_opt}] are plotted as functions
of $J/\gamma$. The strong antibunching can be obtained even if $U_2
< \gamma$, provided $J > \gamma/\sqrt{2}$.
Remarkably, the required
nonlinearity decreases with increasing tunnel coupling  $J$ obeying
Eq.~\eqref{eq:U_opt_approx}.

In Fig.~\ref{fig:2}(b), we show a sketch of the quantum interference
effect responsible for  this counter-intuitive photon antibunching.
The interference is between the following two paths: (a) the direct
excitation from $\ket{10} \xrightarrow{F} \ket{20}$ (solid arrow)
and (b) tunnel-coupling-mediated transition $\ket{10}
\overset{J}{\leftrightarrow} \ket{01} \xrightarrow{F} (\ket{11}
\overset{J}{\leftrightarrow} \ket{02}) \xrightarrow{J} \ket{20}$
(dotted arrows). In order to show in detail the origin of the quantum
interference, we rewrite Eqs.~\eqref{eq:for_C20=0} for $C_{20} = 0$ as
follows. First, we calculate $C_{11}$ from Eqs.~\eqref{eq:C01=C10}
and \eqref{eq:two-particle} neglecting $C_{20}$ as
\begin{align}
C_{11} & = -2JFC_{10}(\DE+U_2-\ii\gamma/2)(\DE-\ii\gamma/2)^{-1}
\nonumber \\ & \quad
[2J^2-4\DE(\DE+U_2)+\gamma^2+\ii2\gamma(2\DE+U_2)]^{-1}.
\end{align}
This amplitude is the result of excitation from $\ket{01}$ to
$\ket{11}$ and of the coupling between $\ket{10}$ and $\ket{01}$ and
also between $\ket{11}$ and $\ket{02}$. From this amplitude,
$C_{20}$ is determined by Eq.~\eqref{C20=C11+C10} as $C_{20}
\propto J C_{11} + F C_{10}$, and we can derive
Eqs.~\eqref{eq:for_C20=0} by the condition $C_{20}=0$.

\begin{figure}[tbp] 
\includegraphics[width=\hsize]{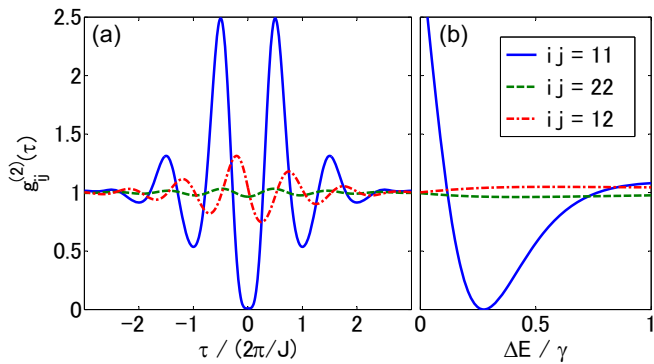}
\caption{(a) The time evolution of the second-order correlation function,
which oscillates with period $2\pi/J$ as the result of amplitude oscillation
between $\ket{01}$ and $\ket{10}$.
(b) Equal-time second-order correlation functions are plotted as functions of
$\Delta E_1 = \Delta E_2 = \Delta E$ normalized to $\gamma_1 = \gamma_2 = \gamma$.
The spectral width of the antibunching resonance is  $\approx 0.3\gamma$.
Parameters: $J = 3\gamma$, $U_1 = U_2 = 0.0428\gamma$, and $F = 0.01\gamma$.
$\Delta E = 0.275\gamma$ in panel (a).}
\label{fig:3}
\end{figure}

As seen in Fig.~\ref{fig:1}(b), while no more than one photon is
present in the first cavity mode at the optimal condition, there can
be more than one photons in the whole system. While there is nearly
perfect antibunching in the driven mode [$g^{(2)}_{11}(\tau=0) <<
1$], the cross-correlation between the two modes exhibits bunching
[$g^{(2)}_{12}(\tau=0) > 1$]. The amplitude oscillation between
$\ket{10}$ and $\ket{01}$ produces the time oscillation of
$g^{(2)}_{11}(\tau)$ with period $2\pi/J$ as reported in
Ref.~\cite{Liew2010PRL} and shown in Fig.~\ref{fig:3}(a).

The equal-time correlation functions is plotted in Fig.~\ref{fig:3}(b) as a
function of the pump detuning $\Delta E/\gamma$: while the optimal
value of the detuning is at $\Delta E = 0.275\gamma$, a strong
antibunching is obtained in a range of about $0.3 \gamma$ around the
optimal value and the width of this window does not significantly
depend on $J/\gamma$. This may suggest that pump pulses of duration
$\Delta t_p$ longer than $1/(0.3\gamma)$ could be enough to ensure
strong antibunching. However, the timescale over which strong
quantum correlations between the photons exist is on the order of
$1/J < \sqrt{2}/\gamma$, as seen in Fig.~\ref{fig:3}(a). While weak
nonlinearities do lead to strong quantum correlations, these
correlations last for a timescale that scales with $1/J\propto\sqrt{U_{\text{opt}}}$ (see
Eq.~\eqref{eq:U_opt_approx}). From a practical perspective, a principal difficulty with
the observation of the photon antibunching with weak nonlinearities
is that it requires fast single-photon detectors \footnote{In a
single nonlinear cavity, the requirement for fast photon detection
can be avoided by using pulsed-laser excitation; this approach does
not work in the system we analyze due to small bandwidth of the
nonlinearity, i.e., the available time of the antibunching is
shorter than the temporal width of pump pulse ($1/J <
\Delta t_{\text{p}}$)}. Conversely, for a given detection set-up, the
required minimal value of the nonlinearity is ultimately determined
by the time resolution of the available single photon detector.

\begin{figure}[tbp] 
\includegraphics[width=\hsize]{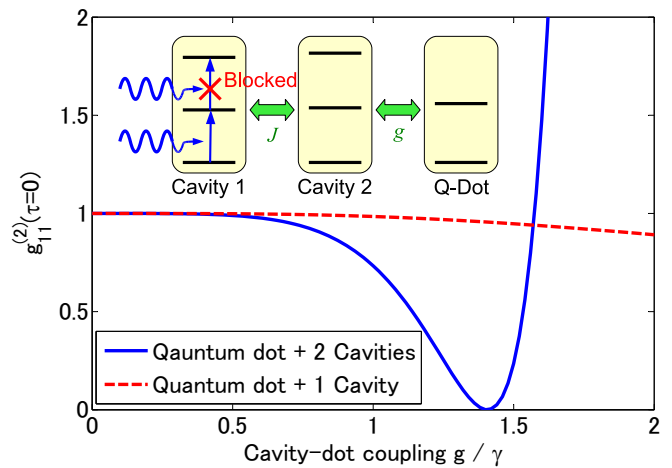}
\caption{Equal-time correlation functions are plotted as functions of coupling strength $g$
between a cavity and a quantum dot.
Solid line represents the results in the system sketched in the inset
[Eq.~\eqref{eq:eq:H_cav-JC}].
Parameters: $\gamma_1 = \gamma_2 = \gamma_{\text{ex}} = \gamma$,
$J = 3\gamma$, $E_1 = \hbar\omega_{\text{p}} + 0.275\gamma$,
$E_2 - E_1 = \gamma$, $E_{\text{ex}} - E_2 = 2\gamma$
and $F = 0.01\gamma$.
Dashed line represents the result in the system with one quantum dot and one cavity
[Jaynes-Cummings model].
$\gamma_1 = \gamma_{\text{ex}} = \gamma$, $E_{\text{ex}} - E_1 = 2\gamma$,
$F = 0.01\gamma$,
and $\hbar\omega_{\text{p}}$ is tuned to the lower one-particle eigenenergy of the Jaynes-Cummings ladder.}
\label{fig:4}
\end{figure}

As seen in Eq.~\eqref{eq:for_C20=0}, the nonlinearity $U_1$ of the
pumped cavity mode is not essential for the antibunching. This means
that only the auxiliary (undriven) photonic mode must have a (weak)
nonlinearity to achieve the quantum interference leading to perfect
photon antibunching. As a practical realization, one could consider
two coupled photonic crystal nanocavities, where the auxiliary
cavity contains a single quantum dot that leads to the required weak
nonlinearity (see the inset in Fig.~\ref{fig:4}). The Hamiltonian is
written as
\begin{align} \label{eq:eq:H_cav-JC} 
\oH_{\text{cav-JC}}
& = \sum_{i=1}^2 E_i \oad_i \oa_i + J(\oad_1\oa_2 + \oad_2\oa_1)
\nonumber \\ & \quad
  + E_{\text{ex}}\ket{\text{ex}}\bra{\text{ex}}
  + g \left( \oad_2\ket{\text{g}}\bra{\text{ex}} + \text{H.c.} \right)
\nonumber \\ & \quad
  + F \ee^{-\ii\wp t}\oad_1 + F^*\ee^{\ii\wp t}\oa_1.
\end{align}
Here, $\ket{\text{g}}$ and $\ket{\text{ex}}$ represent the ground
and excited states of the quantum dot, respectively, $E_{\text{ex}}$
is the excitation energy, and $g$ is the coupling energy with cavity
mode 2. Since the required nonlinearity is relatively weak, one can
use a quantum dot which is off-resonant with respect to the cavity
mode ($|E_{\text{ex}}-E_2|>\gamma_2 = \gamma$) and/or does not
satisfy strong coupling condition ($g \simeq \gamma$). We take the
quantum dot exciton broadening to be equal to the cavity decay rate
for simplicity. We have solved numerically the master equation
associated to the Hamiltonian in Eq.~\eqref{eq:eq:H_cav-JC}.
Fig.~\ref{fig:4} shows $g^{(2)}_{11}(\tau=0)$ of the pumped mode as
a function of $g/\gamma$. The coupling energy between the two
cavities is $J = 3\gamma$, and then the required nonlinear energy
should be $U_{\text{opt}} = 0.0428\gamma$ from Fig.~\ref{fig:2}. In
the present system, this nonlinear energy is practically achieved at
$g = 1.4\gamma$, which is an intermediate strength between the weak-
and strong-coupling regime of cavity mode and quantum dot
excitation. The dashed line in Fig.~\ref{fig:4} represents the
results in the system consisting of one quantum dot and one cavity:
in this ordinary Jaynes-Cummings system, only a small antibunching
is obtained at $g\simeq\gamma$, and the strong-coupling $g\gg\gamma$
is required for the observation of large photon antibunching
\cite{Imamoglu1997PRL,Birnbaum2005N}. In contrast, in the new scheme
using the quantum interference, a nearly perfect antibunching can be
obtained even for $g \simeq \gamma$.

\begin{figure}[tbp] 
\includegraphics[width=\hsize]{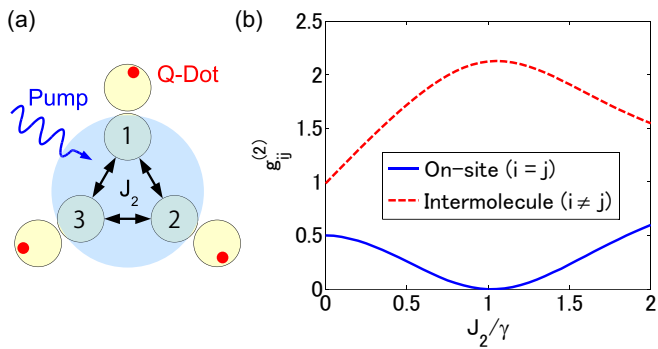}
\caption{(a) Sketch of a triangular lattice of coupled photonic `molecules'.
The driven cavities ($i=1$, 2, and 3) are coupled
with strength $J_2$. (b) The equal-time second-order correlation functions in each mode
(solid line) and between neighbors (dashed line)
are plotted versus $J_2/\gamma$. Parameters:
$J = 3\gamma$, $E_1 = \hbar\omega_{\text{p}} + 0.450\gamma$, $U = 0.0769\gamma$
and $F = 0.01\gamma$.
}
\label{fig:5}
\end{figure}

Finally, we note that the quantum interference can be generalized to
a system of many coupled photonic molecules: in this case, the
strong on-site antibunching can show an interesting interplay with
quantum correlation between neighboring photonic modes. As a
demonstration, we consider a ring of three molecules whose driven
dots are coupled with each other by a tunnel coupling of amplitude
$J_2$ [see Fig.~\ref{fig:5}(a)]. Also in this case a nearly perfect
antibunching can be observed in each driven mode, as shown in the
plots of $g^{(2)}_{ii}(\tau = 0)$ as a function of $J_2/\gamma$ that are
shown as a solid line in Fig.~\ref{fig:5}(b). In order to optimize the
antibunching at a finite value of $J_2\simeq \gamma$, values of $U =
0.0769\gamma$ and $\Delta E = 0.450\gamma$ slightly different from
the single-molecule optimal ones ($U_{\text{opt}}=0.0428\gamma$ and $\Delta
E_{\text{opt}} = 0.275\gamma$) had to be chosen. At the same time, a strong
bunching effect is observed in the equal-time cross-correlation
function between neighboring cavities, which shows a value of
$g^{(2)}_{i\neq j}(0)$ significantly larger than the coherent field value
of $g^{(2)}_{i\neq j}(0)=1$. This remarkable combination of strong
on-site antibunching and strong inter-site bunching suggests that
this system may be a viable alternative to the realization of a
Tonks-Girardeau gas of fermionized photons discussed in Ref.~\cite{Carusotto2009PRL}.

In summary, we have analytically determined that a destructive
quantum interference mechanism is responsible for strong
antibunching in a system consisting of two coupled photonic modes
with small nonlinearity ($U < \gamma$). The quantum interference
effect occurs for an optimal on-site nonlinearity $U_{\text{opt}}
\simeq \frac{2}{3\sqrt{3}}\frac{\gamma^3}{J^2}$, where $J$ is the
intermode tunnel coupling energy and $\gamma$ is the mode
broadening. This robust quantum interference effect has the peculiar
feature that the resulting quantum correlation between the generated
photons survive for timescales much shorter than the photon
lifetime. Nonetheless, we have shown that this quantum interference
scheme has the potential to generate strongly correlated photon
states in arrays of weakly nonlinear cavities.

\end{document}